\documentclass[letterpaper,twocolumn,english,showpacs,preprintnumbers,amsmath,amssymb]{revtex4-1}
\usepackage{amsmath}
\usepackage{amssymb}
\usepackage{graphicx}
\usepackage{esint}

\makeatletter

\pdfpageheight\paperheight
\pdfpagewidth\paperwidth

\@ifundefined{textcolor}{}
{%
 \definecolor{BLACK}{gray}{0}
 \definecolor{WHITE}{gray}{1}
 \definecolor{RED}{rgb}{1,0,0}
 \definecolor{GREEN}{rgb}{0,1,0}
 \definecolor{BLUE}{rgb}{0,0,1}
 \definecolor{CYAN}{cmyk}{1,0,0,0}
 \definecolor{MAGENTA}{cmyk}{0,1,0,0}
 \definecolor{YELLOW}{cmyk}{0,0,1,0}
}

\usepackage{epsfig}%
\usepackage{subfigure}%
\usepackage{dcolumn}%
\usepackage{stmaryrd}%
\usepackage{mathrsfs}%
\usepackage{pifont}%
\usepackage{amsthm}%
\usepackage{bm}%
\usepackage{latexsym}%
\usepackage{amsfonts}%
\newcommand{\beq}{\begin{equation}}
\newcommand{\eeq}{\end{equation}}
\newcommand{\beqa}{\begin{eqnarray}}
\newcommand{\eeqa}{\end{eqnarray}}

\makeatother

\begin{document}

\title{Exact non-Markovian master equations for multiple qubit systems: quantum trajectory approach}

\author{Yusui Chen$^1$\footnote{Email:ychen21@stevens.edu}}

\author{J. Q.  You$^2$\footnote{Email: jqyou@csrc.ac.cn}}

\author{Ting Yu$^1$\footnote{Email:tyu1@stevens.edu}}

\affiliation{$^1$Center for Controlled Quantum Systems and the Department of Physics
and Engineering Physics, Stevens Institute of Technology, Hoboken,
New Jersey 07030, USA\\
$^2$Laboratory for Quantum Optics and Quantum Information, Beijing
Computational Science Research Center, Beijing 100084, China}

\date{\today}

\begin{abstract}
A wide class of exact master equations for a multiple qubit system can be explicitly
constructed by using the corresponding exact non-Markovian quantum state diffusion equations.
These exact master equations arise naturally from the quantum decoherence dynamics
of qubit system as a quantum memory coupled to a collective colored
noisy source. The exact master equations are also important in optimal quantum control, quantum dissipation
and quantum thermodynamics.  In this paper, we show that the exact non-Markovian master equation
for a dissipative N-qubit system can be derived explicitly from the statistical average of
the corresponding non-Markovian quantum trajectories.  We illustrated our general formulation by an explicit construction of
a three-qubit system coupled to a non-Markovian bosonic environment.
This multiple qubit master equation offers an accurate time evolution of quantum systems in various domains, and paves a
way to investigate the memory effect of an open system in a non-Markovian regime without any approximation.
\end{abstract}


\maketitle

\section{Introduction}

A quantum open system, its temporal evolution governed by a master equation or a stochastic Schr\"odinger equation, has attracted a wide-spread interest due to its applications in various research fields such as non-equilibrium quantum dynamics, quantum control, quantum cooling, quantum decoherence and quantum dissipation \cite{Hubook,qc1,qc2,qc3,qc4,qcool1,qcool2,qd1,qd2,qd3,qdiss2,QOS1}.
The quantum dynamics of an open system is commonly formulated in the system plus environment framework where the state of the open system is described by
a reduced density operator. Typically, deriving  the master equation governing the reduced density operator involves several important elements
regarding fine details of the environment and the coupling between the system and environment. In the conventional quantum optics where the quantized radiation field
is treated as an environment,  the master equation for an atomic system weakly coupled to the radiation field is systematically derived, which applies Markov approximation and takes the standard
Lindblad form (setting $\hbar =1$) \cite{QOS2},
\beq
\dot{\rho}=-i[H_{\rm s}, \rho]+\sum_i(2L_i \rho L_i^\dag-L_i^\dag L_i \rho-\rho L_i^\dag L_i).
\eeq
Here, $H_{\rm s}$ is the Hamiltonian of the system of interest, and $L_i$ are a set of system operators called  Lindblad operators
which couple the system to the environment.

An environment can bring about various physical phenomena to the open quantum system \cite{QOS1,QOS3}.
For example, in the case of  two-qubit system coupled to two local bosonic baths,
a Markov environment typically induces both irreversible decoherence and disentanglement
\cite{M1,M2}.  However, the non-Markovian environment with a finite memory time can
assist to regenerate quantum coherence and entanglement in the system \cite{M3,M4,M5,M6}.  Some interesting physics
induced by a non-Markovian environment has been studied extensively by employing
an exact or an approximate non-Markovian master equation  which has many
 experimental applications in  quantum device,  quantum information and quantum optics \cite{NM1,NM2,NM3,NM4,NM5}.

In the case of non-Markovian open system, deriving a non-Markovian master equation is
a notoriously difficult problem due to the lack of a systematic tool that is applicable
to a generic open quantum system irrespective of the system-environment coupling strength
and the environment frequency distribution. For a quantum system coupled to a bosonic or fermionic bath,
a  systematic method is formulated called non-Markovian quantum-state-diffusion method (QSD) or stochastic
Schr\"{o}dinger equation approach \cite{QSD1,QSD2,QSD3}.
In the non-Markovian QSD method, the quantum dynamics represented by a stochastic differential
equation is driven by a Gaussian type of process $z_{t}^{*}$.  By construction,  applying ensemble average on all possible stochastic processes, one
can get the reduced density matrix of interested system. For many
models, such as multilevel atom and multiple qubit system, the exact
non-Markovian dynamics have been numerically studied by using the non-Markovian QSD approach \cite{QSD5,QSD6,QSD7,QSD8,QSD9}.

 It is known that the Markov master equation for the
open system may be derived from the corresponding stochastic unrevealing
\cite{QOS3,MMEQ1,MMEQ2,MMEQ3}. There are also some examples in the non-Markovian
case where the exact master equation can be recovered from the non-Markovian QSD equation, but we need to point out that these works are derived in special conditions, such as single-spin system \cite{QSD5,QSD7,QSD9} or quantum Brownian motion \cite{NMEQ1, NMEQ2}.
For a multiple qubit system, deriving exact master equations from the stochastic
Schr\"{o}dinger equation is still an open problem.  In this paper, we present, for
the first time, a generic exact master equation for multiple qubit dissipative system
coupled to a non-Markovian bosonic bath.  The methodology used in this paper
can also be extended to a multilevel atomic system coupled to a quantized radiation field \cite{Chen-Yu2013b}.  Our exact master equation provides
a systematic tool in dealing with quantum coherence and optimal quantum control in a non-Markovian regime \cite{NMEQ3,NMEQ4}.

Our paper is organized as follows. In Sec.~\ref{ch2},
we introduce a three-qubit system and show the principle idea and the
detail of analytical derivation of exact master equation for the three-qubit system. In Sec.
\ref{ch3}, we show some numerical simulation results by applying the new master equation approach. In Sec.~\ref{ch4},
we start a general discussion on the derivation of the master equation
for N-qubit system.

\section{Exact non-Markovian master equation\label{ch2}}

An N-qubit system representing a carrier of quantum information or memory
is assumed to be coupled to one or more dissipative environments described by a set of harmonic oscillators.
To be specific,  now we consider a three-qubit model to illustrate our method of deriving the exact master equation
from the non-Markovian QSD equation for a multiple qubit system. More generic N-qubit model can be treated in a
similar way.  The total Hamiltonian for our three-qubit system coupled to a bosonic bath may be written as \cite{qd1},
\begin{align}
H_{{\rm tot}} & =H_{{\rm s}}+H_{{\rm int}}+H_{{\rm b}},\label{eq:Htotal}\\
H_{{\rm s}} & =\sum_{j=1}^{3}\frac{\omega_{j}}{2}\sigma_{z}^{j}+J_{xy}\sum_{j=1}^{2}\left(\sigma_{j}^{x}\sigma_{j+1}^{x}+\sigma_{j}^{y}\sigma_{j+1}^{y}\right),\nonumber \\
H_{{\rm int}} & =L\sum_{k}g_{k}b_{k}^{\dagger}+L^{\dagger}\sum_{k}g_{k}b_{k},\nonumber \\
H_{{\rm b}} & =\sum_{k}\omega_{k}b_{k}^{\dagger}b_{k},\nonumber
\end{align}
where $L=\kappa_{1}\sigma_{-}^{1}+\kappa_{2}\sigma_{-}^{2}+\kappa_{3}\sigma_{-}^{3}$,
is the Lindblad operator coupling system to its environment,  $\sigma_{\pm }=(\sigma_{x}\pm i\sigma_{y})/2$ are the
creation (annihilation) operators for a qubit respectively, and $b_{k}(b_{k}^{\dagger})$ is the annihilation (creation) operator
of $k^{th}$ mode in the bosonic environment. Note that $g_{k}$ are the coupling
constants between the system and its environment modes.  For the case of zero temperature environment, the
correlation function for the non-Markovian environment is given by $\alpha(t,s)=\sum_k |g_k|^2 e^{-i\omega_k(t-s)}$.

\subsection{Non-Markovian QSD equation}

The non-Markovian diffusive stochastic Schr\"odinger equation is given by \cite{QSD2}
\begin{align}
\partial_{t}\psi_{t}(z^{*}) & =\left(-iH_{{\rm s}}+Lz_{t}^{*}\right)\psi_{t}(z^{*})\nonumber \\
 & -L^{\dagger}\int_{0}^{t}{\rm d}s\alpha(t,s)\frac{\delta}{\delta z_{s}^{*}}\psi_{t}(z^{*}),\label{eq:QSDeq}
\end{align}
where $\psi_{t}(z^{*})$ is the pure stochastic wave function of the three-qubit system, and $z_{t}^{*}=-i\sum_{k}g_{k}z_{k}^{*}e^{i\omega_{k}t}$
is the complex Gaussian stochastic process with zero mean $\mathcal{M}[z_{t}^{*}]=0$,
and correlations $\mathcal{M}[z_{t}^{*}z_{s}^{*}]=0$ and $\mathcal{M}[z_{t}^{*}z_{s}]=\alpha(t,s)$.
Note that $\alpha(t,s)$ is the correlation
function of the bath, which determines  environment memory time and dictates the transition
from non-Markovian to Markov regimes.
The symbol $\mathcal{M}[...]=\int\frac{{\rm d}^{2}z}{\pi}e^{-|z|^{2}}...$
means ensemble average operation on all stochastic trajectories $z_{t}^{*}$.

The stochastic Schr\"odinger equation (\ref{eq:QSDeq}) can be transformed into a time-local form
when the functional derivative of noise is replaced by an operator $O(t,s,z^{*})=\frac{\delta}{\delta z_{s}^{*}}\psi_{t}(z^{*})$
acting on system's current state. In the Markov limit $O$ operator
must be the same as Lindblad operator $L$, therefore, the consistent initial condition for
$O$ operator is $O(t,t,z^{*})=L$.
By consistency condition $\frac{\partial}{\partial t}\frac{\delta}{\delta z_{s}^{*}}\psi_{t}=\frac{\delta}{\delta z_{s}^{*}}\frac{\partial}{\partial t}\psi_{t}$,
$O$ operator satisfies the following time evolution equation,
\begin{align}
\partial_{t}O(t,s,z^{*}) & =\left[-iH_{{\rm s}}+Lz_{t}^{*}-L^{\dagger}\bar{O}(t,z^{*}),\, O(t,s,z^{*})\right]\nonumber \\
 & -L^{\dagger}\frac{\delta\bar{O}(t,z^{*})}{\delta z_{s}^{*}},\label{eq:consistency}
\end{align}
where $\bar{O}(t,z^{*})=\int_{0}^{t}{\rm d}s\alpha(t,s)O(t,s,z^{*})$.

For the three-qubit system with dissipative coupling, the functional expansion of the $O$ operator contains at most
the two-fold noises \cite{QSD9},
\begin{align}
O(t,s, z^*) & =O_{0}(t,s)+\int_{0}^{t}{\rm d}s_{1}z_{s_{1}}^{*}O_{1}(t,s,s_{1})\nonumber \\
 & +\iint_{0}^{t}{\rm d}s_{1}{\rm d}s_{2}z_{s_{1}}^{*}z_{s_{2}}^{*}O_{2}(t,s,s_{1},s_{2}),\label{eq:O}
\end{align}
where $O_0(t,s), O_1(t,s,s_1), O_2(t,s,s_1,s_2)$ are three $8 \times 8$ matrices, not containing any noise. One can get the evolution
equations for $O_i$ by plugging the solution \eqref{eq:O} into Eq.~(\ref{eq:consistency}),
\begin{align}
 & \left[-iH_{{\rm s}}+Lz_{t}^{*}-L^{\dagger}\bar{O}(t,z^{*}),\, O(t,s,z^{*})\right]-L^{\dagger}\frac{\delta\bar{O}(t,z^*)}{\delta z_{s}^{*}}\nonumber \\
= & \partial_{t}O_{0}(t,s)+\partial_{t}\int_{0}^{t}{\rm d}s_{1}z_{s_{1}}^{*}O_{1}(t,s,s_{1})\nonumber\\
 & +\partial_{t}\iint_{0}^{t}{\rm d}s_{1}{\rm d}s_{2}z_{s_{1}}^{*}z_{s_{2}}^{*}O_{2}(t,s,s_{1},s_{2}).\label{eq:consistency2}
\end{align}
By equating the terms with the same order of noises for two sides of  Eq.~(\ref{eq:consistency2}), a group of differential equations for
$O_{0}$, $O_{1}$ and $O_{2}$  are given by,
\begin{align}
\partial_{t}O_{0}(t,s) & =\left[-iH_{{\rm s}},\: O_{0}\right]-\left[L^{\dagger}\bar{O}_{0},\, O_{0}\right]+L^{\dagger}\bar{O}_{1}(t,s),\nonumber\\
\partial_{t}O_{1}(t,s,s_{1}) & =\left[-iH_{{\rm s}},\: O_{1}\right]-\left[L^{\dagger}\bar{O}_{0},\, O_{1}\right]-\left[L^{\dagger}\bar{O}_{1},\, O_{0}\right]\nonumber \\
 & -L^{\dagger}\left[\bar{O}_{2}(t,s,s_{1})+\bar{O}_{2}(t,s_{1},s)\right],\nonumber \\
\partial_{t}O_{2}(t,s,s_{1},s_{2}) & =\left[-iH_{{\rm s}},\: O_{2}\right]-\left[L^{\dagger}\bar{O}_{0},\, O_{2}\right]-\left[L^{\dagger}\bar{O}_{1},\, O_{1}\right]\nonumber \\
 & -\left[L^{\dagger}\bar{O}_{2},\, O_{0}\right].\label{eq:Opde}
\end{align}
Meanwhile we collect the initial conditions for these operators,
\begin{align*}
O_{1}(t,s,t) &=& \left[L,\, O_{0}(t,s)\right],\\
O_{2}(t,s,s_{1},t)+O_{2}(t,s,t,s_{1}) &=& \left[L,\, O_{1}(t,s,s_{1})\right].
\end{align*}
With the exact $O$ operator, the time-local QSD equation is explicitly determined.
Below, we will show that an exact master equation can be derived from the exact non-Markovian
QSD equation.

\subsection{Formal non-Markovian master equation}

Our approach to the handling of non-Markovian open systems defined by (\ref{eq:Htotal}) aims to deriving the exact non-Markovian master
equation from the corresponding linear QSD equation \eqref{eq:QSDeq}. By design, the reduced density matrix $\rho_{t}$ may be
recovered from the ensemble average over all quantum trajectories. As such, the formal non-Markovian master equation can be written as,
\begin{align*}
{\rm \partial}_{t}\rho_{t} & =-i\left[H_{{\rm s}},\,\rho_{t}\right]+L\mathcal{M}[z_{t}^{*}P_{t}]-L^{\dagger}\mathcal{M}[\bar{O}P_{t}]\\
 & +\mathcal{M}[z_{t}P_{t}]L^{\dagger}-\mathcal{M}[P_{t}\bar{O}^{\dagger}]L,
\end{align*}
where $\rho_{t}=\mathcal{M}[P_{t}]$ and $P_{t}=|\psi_{t}(z^{*})\rangle\langle\psi_{t}(z)|$.
Applying Novikov-type theorem $\mathcal{M}[z_{t}P_{t}]=\int_{0}^{t}{\rm d}s\alpha(t,s)\mathcal{M}[O(t,s,z_t^{*})P_{t}]$,
the above formal master equation can be reorganized as the following compact form,
\begin{equation}
{\rm \partial}_{t}\rho_{t}=-i\left[H_{{\rm s}},\,\rho_{t}\right]+\left[L,\,\mathcal{M}[P_{t}\bar{O}^{\dagger}]\right]-\left[L^{\dagger},\,\mathcal{M}[\bar{O}P_{t}]\right].\label{eq:fromalmeq}
\end{equation}

From Eq.~\eqref{eq:fromalmeq}, it is clear that a corresponding non-Markovian equation may be obtained if one can deal with the terms containing
the ensemble average $\mathcal{M}[\bar{O}P_{t}]$.  In fact, several exact master equations derived from
quantum trajectories have been worked out, including a single two-level system, harmonic oscillator model etc.
However, up to now, deriving an exact master equation for a N-qubit system from the non-Markovian stochastic differential equation
is an unsolved problem.  The major purpose of this paper is explicitly to
show how to accomplish this goal using the three-qubit model as a typical example.  For the three-qubit model considered in this paper,
\begin{align*}
\mathcal{M}[P_{t}\bar{O}^{\dagger}] & =\rho_{t}\bar{O}_{0}^{\dagger}+\int_{0}^{t}{\rm d}s_{1}\mathcal{M}[z_{s_{1}}P_{t}]\bar{O}_{1}^{\dagger}\\
 & +\iint_{0}^{t}{\rm d}s_{1}{\rm d}s_{2}\mathcal{M}[z_{s_{1}}z_{s_{2}}P_{t}]\bar{O}_{2}^{\dagger}.
\end{align*}
Note that, there are two extra terms containing the ensemble averages over noise. Moreover, when we use Novikov-type theorem for
the operator mean value, we should note that the time variables for the $O$ operator are different from that for the stochastic density matrix
$P_{t}$. Hence, we have
\beqa
\mathcal{M}[z_{s_{1}}P_{t}] & =&\int_{0}^{t}{\rm d}s_{2}\alpha_{1,2}\mathcal{M}[O(t,s_{2})P_{t}],\nonumber\\
\mathcal{M}[z_{s_{1}}z_{s_{2}}P_{t}] & =&\int_{0}^{t}{\rm d}s_{3}\alpha_{1,3}\mathcal{M}[z_{s_{2}}O(t,s_{3})P_{t}]\nonumber\\
 & =&\iint_{0}^{t}{\rm d}s_{3}{\rm d}s_{4}\alpha_{1,3}\alpha_{2,4}\mathcal{M}[O(t,s_{3})O(t,s_{4})P_{t}]\nonumber\\
 & +&\iint_{0}^{t}{\rm d}s_{3}{\rm d}s_{4}\alpha_{1,3}\alpha_{2,4}\mathcal{M}[\frac{\delta O(t,s_{3})}{\delta z_{s_{4}}^{*}}P_{t}],\label{eq:newnovikov}
\eeqa
where $\alpha_{i,j}=\alpha(s_{i},s_{j})$. Clearly, all the terms on the right-hand side of Eq.~(\ref{eq:newnovikov}) still involve the statistical average over
 the noise. How to deal with these noisy terms is a crucial step of deriving the exact master equation for a multiple qubit $O$ operator.

\subsection{The derived exact master equation}
To find the exact form of the term $\mathcal{M}[P_{t}\bar{O}^{\dagger}]$,
we recall the Eq.~\eqref{eq:consistency2}, and take a careful analysis on
the structure of each term in $O$ operator expansion.  In the right side of Eq.~\eqref{eq:consistency2},  the highest order of noise, coming from the term $[L^{\dagger}\bar{O},\,O]$, goes to fourth order. While the order
of noise of the right side is up to the second order. These redundant
terms provide a very important observation, named as ``forbidden
conditions'', which take the following form for the three-qubit model,
\beqa
LO_{2} & =0,\quad OO_{2}=0,\nonumber \\
O_{1}O_{1} & =0,\quad O_{1}O_{0}O_{0}=0.\label{eq:forbidden}
\eeqa
Now we deal with the term$\mathcal{M}[P_{t}\bar{O}^{\dagger}]$,
and it is easy to eliminate several complex terms since they satisfy
the ``forbidden conditions''. Thus the compact results in Eq.~\eqref{eq:newnovikov} are:
\begin{align*}
 & \mathcal{M}[O(t,s_{2})P_{t}]\bar{O}_{1}^{\dagger}=O_{0}(t,s_{2})\rho_{t}\bar{O}_{1}^{\dagger}\\
 & +\iint_{0}^{t}ds_{3}{\rm d}s_{4}\alpha_{3,4}O_{1}(t,s_{2},s_{3})\rho_{t}O_{0}^{\dagger}(t,s_{4})\bar{O}_{1}^{\dagger},
\end{align*}
\[
\mathcal{M}[O(t,s_{3})O(t,s_{4})P_{t}]\bar{O}_{2}^{\dagger}=O_{0}(t,s_{3})O_{0}(t,s_{4})\rho_{t}\bar{O}_{2}^{\dagger}
\]
\[
\mathcal{M}[\frac{\delta O(t,s_{3})}{\delta z_{s_{4}}^{*}}P_{t}]\bar{O}_{2}^{\dagger}=O_{1}(t,s_{3},s_{4})\rho_{t}\bar{O}_{2}^{\dagger}.
\]
With above results, the closed form of ensemble average $\mathcal{M}[P_{t}\bar{O}^{\dagger}]$
can be written explicitly as
\begin{widetext}
\begin{align*}
R(t) & =\mathcal{M}[P_{t}\bar{O}^{\dagger}]\\
 & =\rho_{t}\bar{O}_{0}^{\dagger}+\iint_{0}^{t}{\rm d}s_{1}{\rm d}s_{2}\alpha_{1,2}O_{0}(t,s_{2})\rho_{t}\bar{O}_{1}^{\dagger}(t,s_{1})+\iiiint_{0}^{t}{\rm d}s_{1}...{\rm d}s_{4}\alpha_{1,2}\alpha_{3,4}O_{1}(t,s_{2},s_{3})\rho_{t}O_{0}^{\dagger}(t,s_{4})\bar{O}_{1}^{\dagger}(t,s_{1})\\
 & +\iiiint_{0}^{t}{\rm d}s_{1}...{\rm d}s_{4}\alpha_{1,3}\alpha_{2,4}\left[O_{0}(t,s_{3})O_{0}(t,s_{4})+O_{1}(t,s_{3},s_{4})\right]\rho_{t}\bar{O}_{2}^{\dagger}(t,s_{1},s_{2}).
\end{align*}
\end{widetext}

Finally we find the exact non-Markovian master equation for three-qubit system,
\begin{equation}
{\rm \partial}_{t}\rho_{t}=-i\left[H_{{\rm s}},\,\rho_{t}\right]+\left[L,\, R(t)\right]-\left[L^{\dagger},\, R^{\dagger}(t)\right].\label{eq:exactmeq}
\end{equation}
This exact non-Markovian master equation is the major result of this paper. In the following sections, we will apply our result to several interesting cases
where the non-Markovian dynamics is studied by using our derived exact equation.

\section{Numerical calculations\label{ch3}}
Below we study the non-Markovian quantum dynamics of three-qubit system. For simplicity, we use Ornstein-Uhlenbeck noise depicted by
 the correlation function $\alpha(t,s)=\frac{\gamma}{2}e^{-\gamma|t-s|}$.  Although our master equation is universally valid for an arbitrary
 correction function, the advantage of choosing the Ornstein-Uhlenbeck noise is that we can control the single parameter $\gamma$ to
 recover the Markov limit ($\gamma\rightarrow\infty$) from a non-Markov regime.

\begin{figure}
\includegraphics[scale=0.65]{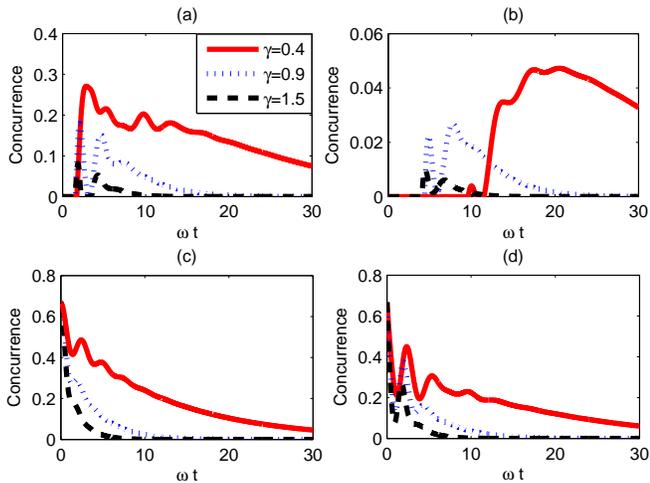}

\caption{(Color online) The dynamics of concurrence between the first and the second qubits
as function of $\omega t$ with different initial states. (a) $|111\rangle$,
(b) $\left(|111\rangle+|000\rangle\right)/\sqrt{2}$, (c) $\left(|100\rangle+|010\rangle+|001\rangle\right)/\sqrt{3}$,
(d) $\left(|110\rangle+|101\rangle+|011\rangle\right)/\sqrt{3}$ .
\label{fig1}}
\end{figure}

In Fig.~\ref{fig1},  four initial states are chosen for three-qubit system.  The entanglement dynamics of selected two qubits (the first
and the second ones) is shown.  Here we choose concurrence as the measurement of entanglement \cite{concurrence}. In Figs.~\ref{fig1} (a) and ~\ref{fig1} (b), initially there is no
entanglement between the two qubits. As the onset of the non-Markovian environment effects,  the generation
of entanglement is observed.  Moreover, as shown in  Fig.~\ref{fig1},  the degree  of the generated entanglement depends sensitively on
 the value of the parameter $\gamma$.  When $\gamma=0.4$ reprinting a longer memory time,  the degree of
 entanglement is almost five times than the case with $\gamma=1.5$ which represents a more Markovian regime.
 In Figs.~\ref{fig1} (c)  and ~\ref{fig1} (d), the initial state of the three-qubit system is the maximally entangled for every two
qubit pair.  When $\gamma=0.4$,  we observe a typical behavior in non-Markovian regime, that is, it exhibits
a stronger entanglement oscillation pattern compared with the case of $\gamma=1.5$.
\begin{figure}
\includegraphics[scale=0.65]{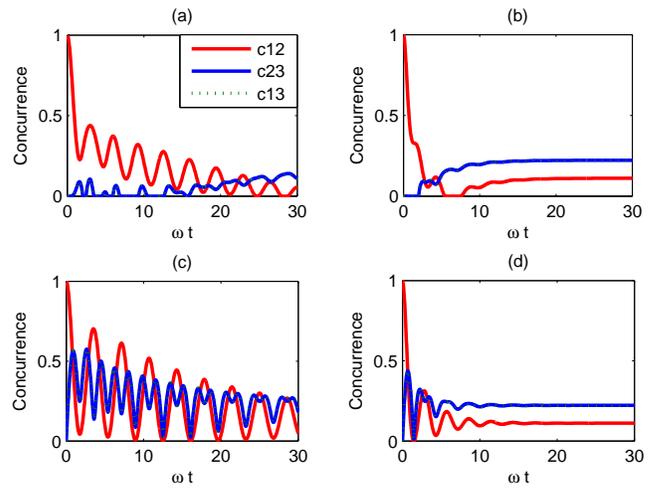}

\caption{(Color online) The dynamics of entanglement flow among three
qubits  $c12$(red solid), $c13$(blue solid) and $c23$(green dotted).
Left column shows a non-Markovian regime with $\gamma=0.4$. Right column
shows a regime close to Markov limit (we choose $\gamma=1.5$).  (a) and (b) use the
same initial state $\left(|11\rangle+|00\rangle\right)\otimes|0\rangle/\sqrt{2}$;
while (c) and (d) use the initial state $\left(|10\rangle+|01\rangle\right)\otimes|0\rangle/\sqrt{2}$.
\label{fig2}}
\end{figure}

In Fig.~\ref{fig2}, the initial state between the first and the second qubits  is Bell state,  and the entanglement flow among three qubits
are studied.  In Figs.~\ref{fig2} (a) and ~\ref{fig2} (c), the system that is coupled to a non-Markovian environment exhibits a strong oscillation,
and shows  a symmetric pattern of the system dynamics.  In particular,  it shows that the entanglement shared by each
pair moves forward and backward between two pairs periodically. In Figs.~\ref{fig2}
(b) and ~\ref{fig2} (d), when $\gamma=1.5$, then the environment is close to the Markov limit,  we see that the state drops
to the final static state quickly as expected for a Markov regime.  It is interesting to note that the entanglement that is present
in one pair initially, will diffuse into three pairs eventually.
As we known, in a two-qubit system coupling to environment, the Bell state $\left(|10\rangle-|01\rangle\right)/\sqrt{2}$ preserves the quantum information. However, in Figs.~\ref{fig2} (b) and ~\ref{fig2} (d), entanglement goes to constant after long time evolution, which shows the direct evidence that the quantum information is easier to be preserved in a three-qubit system than in a two-qubit system. More importantly, the exact master equation for the multiple
qubit systems will allow us systematically study the decoherence issues when the environmental
noises are colored.

\section{General discussions on the exact master equation for an N-qubit system\label{ch4}}

In this section, we give some general discussions on the derivation of non-Markovian master equation
for the N-qubit model, with $H_{{\rm s}}=\sum_{n}\omega_{n}\sigma_{z}^{(n)}$
and $L=\sum_{n}\kappa_{n}\sigma_{-}^{(n)}$. Since we have shown the formal master equation as Eq.~\ref{eq:fromalmeq} which is applicable for general case, so that the goal of deriving exact master equation is based on finding exact $R(t)$. \\
  It is easy to prove that $L^{(N+1)}=0$. Note that, for a matrix polynomial, it means that each term has the same order $N+1$, therefore,
 each term contains at least one zero factor $\left(\sigma_{-}^{(j)}\right)^{2}=0$. On the other hand,
 from Eq.~\eqref{eq:consistency}, we see that the higher order of noise in the new $O$ operator
 come from the commutator relation $z_{t}^{*}[L,\, O]$.
Combining two conditions above, we arrive at our first conclusion that,  for the N-qubit model
considered in this paper,  the highest order of noise is $N-1$, which means that
the $O$ operator contains only finite terms. This is the basic conclusion on which our general discussion is based.
  On the other hand, in Eq.~\eqref{eq:consistency2}, the order of noise
for both sides should match with each other. From the commutator relation
$[L^{\dagger}\bar{O},\, O]$, we have all $O_{j}O_{k}$ terms with
$(j+k)\geq N-1$ would go to zero, since the right side of Eq.~\eqref{eq:consistency2}
doesn't contain such order terms. Now we have a general conclusion for the "forbidden
conditions":
\begin{align}
O_{j}O_{k} & =0,\;\left(j + k \geq N-1\right).\label{eq:generalforbidden}
\end{align}
Generally,  one can get an explicit expression for $\mathcal{M}[P_{t}\bar{O}^{\dagger}]$ after applying Novikov-style theorem for multiple times. We take one term of $R(t)$ as example:
\beqa
& &\mathcal{M}[z_{s_1}...z_{s_{2j-1}}P_{t}\bar{O}_j^{\dagger}]\nonumber\\
&=&\int_0^t\rm{d}s_{2}\alpha_{1,2}\mathcal{M}[z_{s_3}...z_{s_{2j-1}}O^{\dagger}(t,s_2)P_{t}\bar{O}_j^{\dagger}]\nonumber\\
&=&\iint_0^t\rm{d}s_{2}\rm{d}s_{4}\alpha_{1,2}\alpha_{3,4}\mathcal{M}[z_{s_5}...z_{s_{2j-1}}O^{\dagger}(t,s_{2})O^{\dagger}(t,s_{4})P_{t}\bar{O}_j^{\dagger}]\nonumber
\eeqa
where $\alpha_{i,j}=\alpha(s_i,s_j)$. We can keep applying Novikov theorem and do the iteration calculation. With the "forbidden conditions" \eqref{eq:generalforbidden},
it is easy to see that the $\mathcal{M}[P_{t}\bar{O}^{\dagger}]$ will eventually become noise free.
Following this procedure, we can derive the general $R(t)$ and exact master equation for a N-qubit system coupled to a non-Markovian environment.

\section{Conclusion}
We have provided a systematic approach to deriving non-Markovian master equations from the corresponding quantum state diffusion equations.
Non-Markovian master equations for open quantum systems are of  importance in describing quantum dynamics coupled to a non-Markovian
environment.   Our exact non-Markovian master equations provide a new method to handling the
generic open quantum systems where the standard Markov assumption is not longer valid.  For example, our
derived N-qubit master equations would be useful in quantum control and quantum decoherence
of quantum memory and quantum optics where the atomic systems are coupled to a high-Q cavity.  Note that the quantum state diffusion equations
can be formally established for a very wide class of problems in quantum open systems,  therefore, our
method is likely to be useful for many other applications.

\section*{Acknowledgement}
We acknowledge grant support from the NSF PHY-0925174, DOD/AF/AFOSR No. FA9550-12-1-0001 and
the National Basic Research Program of China No. 2014CB921401, the National Natural Science Foundation
of China No. 91121015. TY is grateful to Prof. H. S. Goan for the hospitality during his visit to
the National Taiwan University.

\end{document}